# Temperature fluctuations in H II regions: ionization by Cosmic Rays as a key mechanism

C. Giammanco[1] and J.E. Beckman[1,2]

[1] Instituto de Astrofísica de Canarias, C. Vía Láctea s/n, 38200, La Laguna, Tenerife, Spain
  e-mail: `corrado@ll.iac.es`
[2] Consejo Superior de Investigaciones Científicas, Spain

**Abstract.** We present a detailed model capable of explaining quantitatively the temperature fluctuations observed in luminous, large H II regions. The model is based on two assumptions which we justify on the basis of observations: that the major fraction of the hydrogen in the clouds that form the H II regions is not photoionized and is essentially H I, this H I is lightly ionized by fluxes of low energy cosmic rays (CR) produced by processes originating in the hot stars which illuminate the regions.

**Key words.** ISM: general–ISM: HII regions–ISM: clouds–ISM: photon escape–Elementary particles–Methods: analytical–Methods: data analysis

## 1. Introduction

The problem of the existence of temperature fluctuations in H II regions was first noted by Peimbert (1967), who suggested that these should be measurable in ionized gaseous nebulae. Among modern reasons for thinking that they should be non–zero are the standard photoionization codes, which predict values for the $t^2$ parameter, a measure of these fluctuations which we will define below, of between $10^{-3}$ and $10^{-2}$. However the values measured in H II regions (Stasińska, 2000) are some half an order of magnitude higher than this range. Values of $t^2$ for extragalactic H II regions of between 0.02 and 0.09 are found (González-Delgado et al., 1994, Peimbert et al., 2000, Esteban et al., 2002). These $t^2$ values yield significant effects on element abundance estimates, notably for the primordial $^4$He abundance, $Y_p$, (Steigman et al., 1997, Peimbert et al., 2000). Several authors have modelled mechanisms to explain the origin of these fluctuations, e.g. chemical inhomogeneities (Torres-Peimbert et al., 1990), and the presence of dense gas clumps in the ISM (Viegas & Clegg, 1994). These can explain the $t^2$ values for planetary nebulae, but for H II regions we can quote the verdict of Stasińska (2000) that the fluctuations must be generated by an "unknown physical process".

Esteban (2002), in a comprehensive review of the phenomenon, suggests as a possible origin the shocks caused by stellar winds, as previously studied by Peimbert, Sarmiento and Fierro (1991). However there is considerable evidence that these winds can produce high energy charged particles which can ionize the surrounding medium (Nath & Biermann, 1994, Ramaty, 1996, Webber, 1998). Furthermore a number of authors have shown that there is a major fraction of neutral or lightly ionized gas within H II regions (Yang et al., 1996, Deul and van der Hulst, 1987, Viallefond et al., 1992, Chu and Kennicutt, 1994, Relaño et al., 2005). Based on these two considerations we propose a mechanism which can generate temperature fluctuations in the measured range of amplitudes, and test this via a two phase model.

## 2. The two phase model.

### 2.1. Formulation of the model.

To structure our model we take an H II region to consist of two phases, a warm phase and a cool phase. The warm phase is the gas directly photoionized by stellar radiation, and the cool phase is shielded from the photons, but slightly ionized by charged particles as we will justify in detail below. As the photoionized gas does develop temperature fluctuations, but an order of magnitude smaller than those observed, we will assign a uniform temperature $T_h$ to the warm phase, and a temperature $T_c$ to the cool phase.

As a preliminary step we define, following Peimbert (1967) the mean temperature $T_\circ$ and the squared temperature fluctuation $t^2$ as follows:

$$T_\circ = \frac{\int_V n_e\, n_i\, T_e\, dv}{\int_V n_e\, n_i\, dv}, \qquad (1)$$

$$t^2 = \frac{\int_V n_e\, n_i\, (T_e^2 - T_\circ^2)\, dv}{T_\circ^2 \int_V n_e\, n_i\, dv}. \qquad (2)$$

Using the assumptions of a two phase model the equations become:

$$T_\circ = \frac{\int_{V_h} n_e^h\, n_i^h\, T_h\, dv + \int_{V_c} n_e^c\, n_i^c\, T_c\, dv}{\int_{V_h} n_e^h\, n_i^h\, dv + \int_{V_c} n_e^c\, n_i^c\, dv}, \qquad (3)$$

$$t^2 = \frac{\int_{V_h} n_e^h n_i^h (T_h^2 - T_o^2)\, dv + \int_{V_c} n_e^c n_i^c (T_c^2 - T_o^2)\, dv}{T_o^2 \left(\int_{V_h} n_e^h n_i^h\, dv + \int_{V_c} n_e^c n_i^c\, dv\right)}, \quad (4)$$

where $V_h$, $n_e^h$ and $n_i^h$ are respectively the total volume occupied by the warm phase, the corresponding electron density, and the ion density (in this case that of $H^+$) and $V_c$, $n_e^c$, $n_i^c$ are the corresponding parameters for the cool phase.

To simplify the handling of the equations we define two weights $w_h$ and $w_c$ by:

$$w_c = 1 - w_h, \quad (5)$$

$$w_h = \frac{\int_{V_h} n^2\, dv}{\int_{V_h} n^2\, dv + \int_{V_c} x^2 m^2\, dv}. \quad (6)$$

Here we call $n$ the density of the warm phase, for which we take its ionization fraction as close to unity (see Giammanco et al. 2004), so that $n_e^h n_i^h$ can be replaced by $n^2$. Similarly we term $m$ the density of the cool phase, while $x$ is its ionization fraction.

Se now we find:

$$w_h = \frac{n^2 V_h}{n^2 V_h + x^2 m^2 V_c}, \quad (7)$$

from which we obtain

$$\frac{1}{w_h} - 1 = \frac{x^2 m^2 V_c}{n^2 V_h} = x^2 \frac{m}{n} \theta. \quad (8)$$

We define $\theta$ as the ratio of the cool to warm gas mas $M_c/M_h = mV_c/nV_h$. We will simplify the treatment by taking the limiting case of $m = n$ but will later consider the effect of taking $m \neq n$, since $m$ should in general be larger than $n$. Using the definitions of the weighting factors we rewrite equations (3) and (4) as:

$$T_o = T_h w_h + T_c (1 - w_h), \quad (9)$$

$$t^2 = \frac{T_h^2 w_h + T_c^2 (1 - w_h) - T_o^2}{T_o^2}, \quad (10)$$

from which we derive the following expression for the temperature of the cool phase:

$$T_c = T_o \left(1 - \sqrt{\frac{t^2}{x^2 \theta}}\right), \quad (11)$$

and for that of the warm phase:

$$T_h = T_o \left(1 + \sqrt{t^2 x^2 \theta}\right). \quad (12)$$

### 2.2. Estimates of $x^2 \theta$

We could calculate $x^2 \theta$ using only Eqs. (11) and (12) if we knew either $T_c$ or $T_h$ which are not known *a priori*. However we can set useful constraints based on observations. Examining estimates of mean temperatures observed in representative extragalactic H II regions which range from 8000 K to 16000 K, we find that they are quite close to the values computed using photoionization codes with realistic stellar photoionizing spectra.

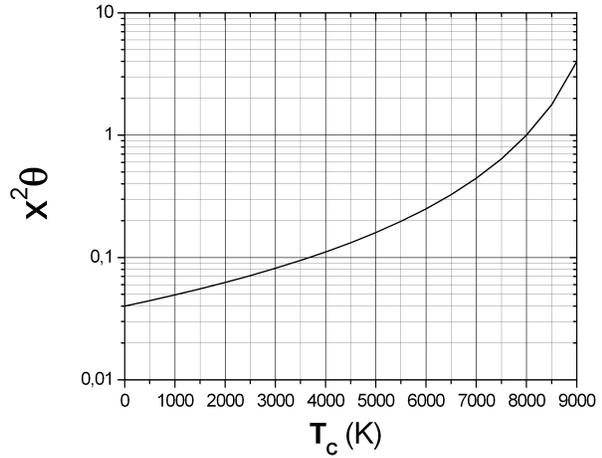

**Fig. 1.** The value of $x^2\theta$ plotted against an assumed variation of $T_c$ for a model with a value of $t^2$ of $4\times10^{-2}$, and $T_o$=10000 K.

Assuming here that the photoionized component is the warm component, this means that the values for $T_h$ are close to those for $T_o$, which from Eq. (12) means that the $x^2\theta$ must be small. From Eq. (11) we find that in the limit as $T_c \to 0$, $x^2\theta \to t^2$. Although values for $T_c$ approaching zero are clearly not realistic, any temperature for the cool phase below say, 500 K yields values for $x^2\theta$ which are quite insensitive to variations in $T_c$ as seen in Fig.1, so that we can take as a practical approximation:

$$x^2\theta \sim t^2. \quad (13)$$

In Table 1 we gives estimates of $x^2\theta$ as functions of the assumed values of $T_c$ for a set of H II regions analysed by Esteban et al. (2002).

## 3. The cool gas

There have been a number of pieces of evidence in the literature pointing to the presence of a large fraction of neutral gas as a component of luminous extragalactic H II regions. In a thorough, detailed study of NGC 604 in M 33, Yang et al. (1996), showed that the width of the observed integrated H$\alpha$ emission line corresponds to that predicted by the virial theorem, if the total gas mass is $\sim$10 times that of the observed ionized gas mass. This measurement served to confirm previous direct measurements by Deul and van der Hulst (1987) and by Viallefond et al. (1992) which indicated that the combined mass of H I and H II is an order of magnitude higher than the mass of H II in the region. Using dynamical arguments, Chu and Kennicutt (1994) estimated a neutral to ionized gas mass ratio of more than 10 for 30 Doradus, and Relaño et al. (2005) used similar arguments to show, for a sample of some tens of the most luminous H II regions in nearby galaxies that their neutral hydrogen components dominate the ionized components in mass by factors varying from 5 to over 20.

The presence of strong density fluctuations in H II regions has long been demonstrated in the discrepancy between *in situ* electron density, $n_e$, estimates using line ratios (e.g. Zaritsky et al., 1994) which yield values for $n_e$ of $\gtrsim 100$ cm$^{-3}$ and estimates of mean electron densities for complete regions using emission measures for single lines, which yield values of $\gtrsim 1$ cm$^{-3}$. This discrepancy is interpreted using a "filling factor" of dense clumps in a tenuous interclump medium, introduced by Osterbrock and Flather (1959). In a study based on these parameters Giammanco et al. (2004) showed that for a simplified model distribution of spherical clumps of ~ 1 pc size in a luminous H II region of radius 100 pc the photoionized portion of a clump (apart from those few within 10 pc of the stellar sources) will be a shallow zone on the starward side of the clump, which effectively shields the rest of the clump leaving it neutral. The zone depth depends on the distance of the clump from the ionizing sources, but an average neutral to ionized gas mass ratio over the whole region of ten or more is characteristic of these models. These estimates depend little on the assumed structure of the dense component, whether clumpy or filamentary. For the purpose of the present model accounting for the temperature fluctuations the geometry of the dense component is not important; all that is required is the presence of a major component of cool gas which is neutral, or has a very low ionization fraction.

| Objeto | n | $T_\circ$ | $t^2$ | $(x^2\theta)_{100}$ | $T_h^{100}$ | $(x^2\theta)_{1000}$ | $T_h^{1000}$ | $(x^2\theta)_{4000}$ | $T_h^{4000}$ |
|---|---|---|---|---|---|---|---|---|---|
| NGC 604 | $\leq 100$ | 8150±150 | 0.027±0.018 | 0.028 | 8372 | 0.035 | 8401 | 0.10 | 8582 |
| NGC 5461 | 300±70 | 8600±250 | 0.041±0.021 | 0.042 | 8957 | 0.052 | 8999 | 0.14 | 9259 |
| NGC 5471 | 220±70 | 14100±300 | 0.074: | 0.075 | 15151 | 0.086 | 15223 | 0.14 | 15557 |
| NGC 2363 | 360±90 | 15700±300 | 0.128±0.045 | 0.130 | 17722 | 0.146 | 17846 | 0.23 | 18397 |

**Table 1.** Estimates of $x^2\theta$ and $T_h$ for a set of H II regions whose parameters are given in Esteban et al. (2002), which is one of the most complete sources of data on $T_\circ$ and $t^2$ for extragalactic H II regions. In the first three data columns we give the electron density of the clumps, the mean temperatures, and the values of $t^2$ taken from that article. In the succeeding columns are our estimates of $x^2\theta$ and $T_h$ assuming values for $T_c$ of 100 K, 1000 K, and 4000 K respectively. One clear result is that for values of $T_c$ below 1000 K the estimates of $x^2\theta$ are very insensitive to the temperature, and are close to the corresponding values of $t^2$.

## 4. The hypothesis quantified.

The first predictions of the ionization of the insterstellar medium (ISM) by cosmic rays date from over 40 years ago (Hayakawa Nishimura & Takayanagi, 1961). Their work was refined by Spitzer & Tomasko (1968), in a classical article in which the following model was presented. Cosmic rays interacting with the neutral ISM can ionize H atoms at a rate $\zeta$ determined by the flux $j(E)\,dE$ of the ionizing particle with energy $E$, and their ionization cross–section $\sigma(E)$, given in a formula due to Bethe (1933, Eq. [55.13]). The number of ionizations per second per unit volume in a medium of density $n$ cm$^{-3}$ due to cosmic rays is given by:

$$N = n\,\frac{5}{3}\int 4\pi\,j(E)\,\sigma(E)\,dE = n\,\zeta, \qquad (14)$$

where the factor 5/3 takes into account the relativistic electrons produced in the ionization, which go on to produce further ionization. The ionization fraction $x$ produced in the medium is given by the equation of detailed balance which, using the "*on the spot*" approximation, can be written:

$$n\,\zeta\,(1-x) - n^2\,x^2\,\beta_2 = 0, \qquad (15)$$

where $\beta_2$ is the recombination coefficient for the hydrogen atom to all levels above the fundamental. The general differential Galactic cosmic ray flux spectrum $j(E)$ can be well measured locally at relatively high energies, showing a maximum value at ~ 1 GeV. However the form of $\sigma(E)$ causes the main contribution to (14) to lie between 100 MeV, for nucleons, and ~10 MeV, for electrons (Webber, 1998). The present canonical value for $\zeta$, given by Black (1990), and by van Dishoek and Black (1991), is 3–7 ×10$^{-17}$ s$^{-1}$. However to this global Galactic flux must be added various more or less strong "anomalous" fluxes, due to localized sources of highly ionizing particles at MeV energies. As early as 1972 Peimbert and Goldsmith (1972) proposed the hypothesis of the existence of a low energy cosmic ray flux in the Orion nebula to explain the intensity of the $\lambda$ 4686 line of He II emitted from the nebula. Subsequently, using the COMPTEL satellite Bloemen et al. (1994) discovered a gamma ray source in Orion and Ramaty (1996) inferred the presence of a strong local cosmic ray flux. He derived for the corresponding ionization rate a value of 0.5–1.0 ×10$^{-13}$ s$^{-1}$ M$_5^{-1}$ where M$_5$ is the irradiated mass of neutral hydrogen in the Orion complex in units of 10$^5$ M$_\odot$. We take from Maddalena et al. (1986) an observed value for this mass as ~ 4–7 ×10$^5$ M$_\odot$, which gives a value for $\zeta$ for this source of 2–7 ×10$^{-13}$ s$^{-1}$.

Based on an extensive bibliographic study Ramaty (1996) proposed a shock acceleration mechnism for the particles linked to O and B star winds, previously proposed by Nath and Biermann (1994). As particle generation mechanisms he suggested collectively the winds of WR stars (Ramaty et al. 1995a) supernova remnants (Cass´e et al. 1995, Ramaty et al. 1996) and the effects of IS grain disintegration by shocks (Ramaty et al. 1995b, 1996), all of which are clearly probable and relate to the stellar content of H II regions. Finally we should mention that Webber (1998) finding an anomalous cosmic ray component accelerated in the termination shock of the heliosphere, and knowing Ramaty's work, suggested that all H II regions ought to be significant sources of low energy cosmic rays.

Now we look at the effect of cosmic ray ionization on the temperature of the ISM. Spitzer and Tomasko (1968) calculated that each ionization event should contribute 34 eV to the gas. Taking into account the relevant cooling rate they found a relation between $\zeta/n$ and the gas temperature $T$. Taking 100 cm$^{-3}$ for $n$ (a lower limit canonical value for the cool "neutral" phase) and using the ionization rate for Orion we find a value for $T$ of ~60 K, and for an ionizing CR flux ten times greater this rises to ~100 K. These values are compatible with the requirements of the model for deriving the observed values of $t^2$.

### 4.1. The relation between $\zeta$ y $t^2$

The final step in our scenario is to calculate, assuming that the cool component is in fact partially ionized by localized cosmic rays, the ionization rate required to yield a given value of $t^2$. Using the low temperature limit:

$$x^2 \sim \frac{t^2}{\theta}, \quad (16)$$

and using equation (15) we have:

$$\zeta \sim \frac{t^2}{\theta} \frac{\beta_2 \, n}{1-x}. \quad (17)$$

We take from Spitzer (1948) the relation between $\beta_2$ and the gas temperature $T$, valid for the full temperature range between 30 K and 1000 K, as

$$\beta_2 = \frac{2.07 \times 10^{-11}}{\sqrt{T}}. \quad (18)$$

When comparing observational results with this theoretical model we must be aware that values of $t^2$ may be derived from observations of more than one type. Specifically, the values shown in Table 1 were obtained using the ion $O^{++}$ while the ionization rate $\zeta$ and the recombination coefficient which we have used are those corresponding to H. To make a more valid comparison we should use values of $t^2$ derived from observations of H emission. In spite of this we first point out that the neutral H mass for NGC 604 is known to be an order of magnitude higher than the ionized mass, so that it is fair to assume that the value of $t^2$ given by Esteban et al. (2002) for this region (see Table 1) based on oxygen emission lines, will be similar, at least in order of magnitude, to the value which would be derived using hydrogen. In that case if we apply the expression in Eq. (17) we obtain $\zeta < 6 \times 10^{-13}$. However we can find a case for which $t^2$ is derived using H emission, as Peimbert (2003) gives a value of $t^2$ for 30 Doradus, obtained using the Balmer continuum, of $t^2 = 0.022$. Taking the value of $n_e$ 316 cm$^{-3}$ from the same article, and applying the same ratio of 10 for the neutral to ionized gas ratio we find $\zeta \sim 2 \times 10^{-12}$, a very reasonable result considering the ratio of luminosities of NGC 604 and 30 Dor.

In our development of the model to this point we have simplified by taking the densities of the cool ($m$) and warm phases ($n$) to be equal. Relaxing this condition, Eq. (16) becomes $x^2 \sim (t^2 \, n)/(\theta \, m)$ and substituting $m$ for $n$ in Eq. (15) yields an invariant form in $n$ for Eq. (17). However $m \geq n$ so the value of $x$ is reduced, and from (17) the value of $\zeta$ is thus reduced, so the required ionization rate can never exceed the value derived using $n = m$.

### 5. Conclusion.

The value for $\zeta$ we have derived to account for the measured value of $t^2$ for NGC 604 is to the high end of the range inferred for the Orion nebula from the gamma–ray observations and for 30 Dordadus the value is again somewhat higher. From the ratio of their H$\alpha$ luminosities, the stellar ionizing luminosities of NGC 604 and 30 Doradus must be well over an order of magnitude higher than that for the Orion nebula (Kennicutt, 1984) and assuming that the rate of CR production is a rising function of the stellar luminosity the values of $\zeta$ for the two objects are clearly compatible with the basic hypothesis that CR ionization of the cool component is a key mechanism for producing the observed temperature fluctuations. There is no reason to believe that either NGC 604 or 30 Doradus are atypical in this respect. We consider that the mechanism should work adequately in any such region.

The formalism presented here, which considers a two phase model, can be readily adapted to the study of alternative scenarios for the production of the measured values of $t^2$.

*Acknowledgements.* We wish to thank Jorge García–Rojas for the discussions about $t^2$, Marco Casolino for discussions about cosmic rays and Héctor Castañeda for discussion about Orion properties. We also thank the referee, Manuel Peimbert for the constructive report. This study was supported by grant AYA 2004-08251-CO2-01, and by the Canary Islands Consejería de Empleo y Asuntos Sociales.